\documentclass[12pt]{article} 
\usepackage[sectionbib]{natbib}
\usepackage{array,epsfig,fancyheadings,rotating}
\usepackage[]{hyperref}  
\usepackage{sectsty, secdot}
\sectionfont{\fontsize{12}{14pt plus.8pt minus .6pt}\selectfont}
\renewcommand{\theequation}{\thesection\arabic{equation}}
\subsectionfont{\fontsize{12}{14pt plus.8pt minus .6pt}\selectfont}

\usepackage{amsmath}
\usepackage{amssymb}
\usepackage{amsfonts}
\usepackage{multirow}
\usepackage{amsthm}
\usepackage{cleveref}

\newtheorem{innercustomgeneric}{\customgenericname}
\providecommand{\customgenericname}{}
\newcommand{\newcustomtheorem}[2]{%
	\newenvironment{#1}[1]
	{%
		\renewcommand\customgenericname{#2}%
		\renewcommand\theinnercustomgeneric{##1}%
		\innercustomgeneric
	}
	{\endinnercustomgeneric}
}

\newcustomtheorem{customcondition}{Condition}


\newcounter{parentnumber}

\newtheorem{theorem}{Theorem}
\newtheorem{lemma}{Lemma}
\newtheorem{corollary}{Corollary}
\newtheorem{proposition}{Proposition}
\newtheorem{condition}{Condition}
\theoremstyle{definition}

\usepackage{mathrsfs}
\usepackage{natbib, graphicx, color, setspace, amssymb, amsmath, amsthm, amstext, booktabs, fancyhdr, multirow, float, bm, lineno, xr, subfigure,enumerate,url}
\usepackage{times}
\allowdisplaybreaks 
\usepackage{pdflscape}
\usepackage{placeins} 
\usepackage{authblk} 
\usepackage{longtable}
\usepackage[ruled,vlined,onelanguage]{algorithm2e}
\usepackage{comment}

\usepackage{xcite}
\makeatletter
\newcommand*{\addFileDependency}[1]{
  \typeout{(#1)}
  \@addtofilelist{#1}
  \IfFileExists{#1}{}{\typeout{No file #1.}}
}
\makeatother

\newcommand*{\myexternaldocument}[1]{%
    \externaldocument{#1}%
    \addFileDependency{#1.tex}%
    \addFileDependency{#1.aux}%
}

\myexternaldocument{Supplementary}

\newcommand*\patchAmsMathEnvironmentForLineno[1]{%
	\expandafter\let\csname old#1\expandafter\endcsname\csname #1\endcsname
	\expandafter\let\csname oldend#1\expandafter\endcsname\csname end#1\endcsname
	\renewenvironment{#1}%
	{\linenomath\csname old#1\endcsname}%
	{\csname oldend#1\endcsname\endlinenomath}}%
\newcommand*\patchBothAmsMathEnvironmentsForLineno[1]{%
	\patchAmsMathEnvironmentForLineno{#1}%
	\patchAmsMathEnvironmentForLineno{#1*}}%
\AtBeginDocument{%
	\patchBothAmsMathEnvironmentsForLineno{equation}%
	\patchBothAmsMathEnvironmentsForLineno{align}%
	\patchBothAmsMathEnvironmentsForLineno{flalign}%
	\patchBothAmsMathEnvironmentsForLineno{alignat}%
	\patchBothAmsMathEnvironmentsForLineno{gather}%
	\patchBothAmsMathEnvironmentsForLineno{multline}%
}
\usepackage{etoolbox}
\usepackage{tikz}
\usepackage{array}

\newcommand{\bmR}{\bm{R}}

\newcommand{\bmalpha}{\bm{\alpha}}
\newcommand{\bmbeta}{\bm{\beta}}
\newcommand{\bmtheta}{\bm{\theta}}
\newcommand{\bmepsilon}{\bm{\epsilon}}
\newcommand{\bmvarepsilon}{\bm{\varepsilon}}
\newcommand{\bmphi}{\bm{\phi}}
\newcommand{\bmrho}{\bm{\rho}}
\newcommand{\bmmu}{\bm{\mu}}
\newcommand{\bmY}{\bm{Y}}
\newcommand{\bmX}{\bm{X}}
\newcommand{\bmW}{\bm{W}}
\newcommand{\bmA}{\bm{A}}
\newcommand{\bmSigma}{\bm{\Sigma}}
\newcommand{\bmXi}{\bm{\Xi}}
\newcommand{\bmthetao}{\bmtheta^{(0)}}
\newcommand{\bmpsi}{\bm{\psi}}

\newcommand{\bmbetao}{\bmbeta^{(0)}}
\newcommand{\bmalphao}{\bmalpha^{(0)}}
\newcommand{\bmrhoo}{\bmrho^{(0)}}
\newcommand{\betao}{\beta^{(0)}}
\newcommand{\rhoo}{\rho^{(0)}}

\newcommand{\bmAhalf}{\bm{A}^{1/2}}
\newcommand{\bmANhalf}{\bm{A}^{-1/2}}

\newcommand{\bmvarthetao}{\bm{\vartheta}^{(0)}}
\newcommand{\bmvartheta}{\bm{\vartheta}}
\newcommand{\tildebmSigma}{\tilde{\bmSigma}}

\newcommand{\barbmXi}{\bar{\bm{\Xi}}}

\DeclareMathAlphabet{\mathbcal}{OMS}{cmsy}{b}{n}

\newcommand{\barmathbcalA}{\bar{\bm{Z}}}
\newcommand{\tbmmA}{\tilde{\bm{Z}}}

\def\beq{\begin{equation}}
\def\eeq{\end{equation}}
\def\beqr{\begin{eqnarray}}
\def\eeqr{\end{eqnarray}}
\def\beqrs{\begin{eqnarray*}}
	\def\eeqrs{\end{eqnarray*}}
\def\bet{\begin{theorem}}
	\def\eet{\end{theorem}}
\def\bel{\begin{lemma}}
	\def\eel{\end{lemma}}
\def\bep{\begin{proposition}}
	\def\eep{\end{proposition}}
\def\bg{\begin{figure}[tbph]\begin{center}}
		\def\eg{\end{center}\end{figure}}

\def\bc{\begin{center}}
	\def\ec{\end{center}}

\def\mS{\mathcal S}

\def\bmmA{\bm{Z}}

\def\be{\begin{equation}}
\def\ee{\end{equation}}
\def\ben{\begin{equation*}}
\def\een{\end{equation*}}
\def\bea{\begin{eqnarray}}
\def\eea{\end{eqnarray}}

\def\bda{\begin{eqnarray*}}
	\def\eda{\end{eqnarray*}}

\def\lsk{\left(}
\def\rsk{\right)}
\def\lbk{\left \{ }
\def\rbk{\right \} }

\def\rmE{E}

\def\tr{\mathrm{tr}}

\begin{document}
	
	
	\renewcommand{\baselinestretch}{2}
	
	\markright{ \hbox{\footnotesize\rm Statistica Sinica
		}\hfill\\[-13pt]
		\hbox{\footnotesize\rm
		}\hfill }
	
	\markboth{\hfill{\footnotesize\rm Zhi Yang Tho AND Francis K.C. Hui AND Tao Zou} \hfill}
	{\hfill {\footnotesize\rm Joint Mean and Correlation Regression Models} \hfill}
	
	\renewcommand{\thefootnote}{}
	$\ $\par
	
	\renewcommand{\thefootnote}{*} 
	\fontsize{12}{14pt plus.8pt minus .6pt}\selectfont \vspace{0.8pc}
	\centerline{\large\bf Joint Mean and Correlation Regression Models}
	\vspace{2pt} 
	\centerline{\large\bf for Multivariate Data}
	\vspace{.4cm} 
	\centerline{Zhi Yang Tho\footnote{Corresponding author.}, Francis K. C. Hui, and Tao Zou\thanks{Corresponding author.}} 
	\vspace{.4cm} 
	\centerline{\it The Australian National University}
	\vspace{.55cm} \fontsize{9}{11.5pt plus.8pt minus.6pt}\selectfont
	
	
	\begin{quotation}
		\noindent {\it Abstract:}
We propose a joint mean and correlation regression model for multivariate discrete and (semi-)continuous response data, that simultaneously regresses the mean of each response against a set of covariates, and the correlations between responses against a set of similarity/distance measures. A set of joint estimating equations are formulated to construct an estimator of both the mean regression coefficients and the correlation regression parameters. Under a general setting where the number of responses can tend to infinity, the joint estimator is demonstrated to be consistent and asymptotically normally distributed, with differing rates of convergence due to the mean regression coefficients being heterogeneous across responses. An iterative estimation procedure is developed to obtain parameter estimates in the required (constrained) parameter space. Simulations demonstrate the strong finite sample performance of the proposed estimator in terms of point estimation and inference. We apply the proposed model to a count dataset of 38 Carabidae ground beetle species sampled throughout Scotland, along with information about the environmental conditions of each site and the traits of each species. Results show the relationship between mean abundance and environmental covariates differs across the beetle species, and that beetle total length is important in driving the correlations between species.

		\vspace{9pt}
		\par
	\end{quotation}\par

	\def\thefigure{\arabic{figure}}
	\def\thetable{\arabic{table}}
	
	\renewcommand{\theequation}{\thesection.\arabic{equation}}

	\fontsize{12}{14pt plus.8pt minus .6pt}\selectfont

\section{Introduction} \label{sec:intro}
The analysis of correlated multivariate or multi-response data is becoming increasingly important nowadays, as it provides enhanced opportunities to answer more diverse and deeper scientific questions relative to studying a univariate response. A prime example is the ability to simultaneously study how responses vary together as a function of covariates, as well as how correlations between responses are related to similarity (or distance) measures of predictors. For instance, in ecology there is a growing interest in jointly modeling how species' distributions are associated with environmental covariates along with how the covariation between species varies with phylogenetic and trait distances \citep{wartonETAL2015,tikhonov2017,ovaskainen2020joint}. 

To study how the means of a set of responses vary as a function of covariates, a popular approach in the statistical literature is generalized estimating equations \citep[GEEs,][]{liangANDzeger1986} or variations thereof, where a mean model for each response is coupled with a working, between-response correlation matrix. The latter can be specified in different ways, leading to varying degrees of improved statistical efficiency for inference on the mean regression coefficients \citep[e.g.,][]{quEtAl2000,yeANDpan2006,warton2011regularized}. GEEs focus primarily on modeling and estimating the mean component; they are not designed to answer scientific questions relating to the correlation between responses themselves. On the other hand, various methods have been developed that explicitly link the covariance matrix of the response vector, or functions thereof, to a linear combination of known symmetric matrices \citep[e.g.,][]{anderson1973,chiuETAL1996,zwiernikETAL2017}. Of particular note are the studies of \cite{pourahmadi1999}, \cite{zhangANDleng2012}, \cite{zhangANDlengANDtang2015} and \cite{bonat2016}, who integrated mean modeling with multivariate covariance modeling using covariate information under specific model structures. \label{page:rewording_start0}These studies focused primarily on longitudinal or spatio-temporal data, so they considered common structures e.g., compound symmetry or neighborhood structure in \cite{bonat2016}, to account for known longitudinal or spatio-temporal associations in the data.\label{page:rewording_end0} \label{page:huETAL_start0}More recently, \cite{huETAL2024} considered to regress the elements of the generalized z-transformed correlation matrices of general correlated data on covariates that are formed by taking the difference of the predictors associated with each pair of responses, while \cite{tangETAL2019} studied a joint mean-correlation regression for discrete longitudinal data by modeling the correlation matrix in a Gaussian copula via a hyperspherical reparameterization.\label{page:huETAL_end0}

In this article, we propose a new joint mean and correlation regression model that simultaneously quantifies the relationship between the mean of each response and a set of covariates, and the relationship between the response correlation matrix and a set of similarity measures of predictors. Analogous to classical GEEs, the proposed joint model requires specifying only the first two moments of each response along with the correlation matrix between responses. The first moment of each response is regressed against a set of covariates with the help of link functions, and using mean regression coefficients that are heterogeneous across responses i.e., each response has its own set of regression coefficients. The specification of the second moment uses variance functions to capture potential mean-variance relationships within each response. \label{page:W_observed1_start}Turning to the between-response correlation matrix, we propose to regress this against a set of similarity matrices that are either observed directly as part of the data collection process, or induced from available predictor information associated with each response.\label{page:W_observed1_end} As one example of such similarity measures, in text frequency analysis where the responses are frequencies of different words across documents, similarity matrices can be formed from various characteristics of the words such as their topical meanings \citep[e.g.,][]{zhuANDxing2011}; see also Section \ref{sec:ground_beetle} for similarity matrices formed from species traits in ecology.
\label{page:rewording_start} Unlike the aforementioned works on joint mean-covariance modeling that focused on longitudinal or spatio-temporal data, we aim to provide an explicit, data-driven quantification of how the unknown dependence structure for general correlated data is informed by these similarity measures of auxiliary predictor information, through a set of correlation regression parameters that are common across responses.\label{page:rewording_end}\label{page:huETAL_start} Furthermore, unlike the studies of \cite{tangETAL2019} and \cite{huETAL2024} that considered hyperspherical reparameterization and generalized z-transformation of the correlation matrices, respectively, we directly model the correlation matrix, noting that the covariates constructed by \cite{huETAL2024} are analogous to our idea of similarity measures.
\label{page:huETAL_end}

The idea of correlation regression builds upon recent developments of covariance regression in \cite{tao2017,tao2020,ZOU2021}, although we make three clear advances on these works. First, we extend covariance regression to correlation regression, while allowing for heterogeneous variances across responses. This differs from \cite{tao2017,tao2020,ZOU2021} who all assumed a homogeneous variance across responses. \label{page:GEE_framework1_start}Note also that we model the correlation rather than the covariance matrix, as our method is developed within the GEE framework which requires the specification of a working correlation matrix.\label{page:GEE_framework1_end} Second, our proposed joint mean and correlation regression model can handle a much wider variety of discrete and (semi-)continuous responses e.g., overdispersed counts, binomial, and non-negative continuous responses that arise in ecology and quantitative genetics among other fields. Third, we develop and study asymptotic inference for both the heterogeneous mean regression coefficients and correlation regression parameters, in contrast to \cite{tao2017,tao2020,ZOU2021} who focused solely on the covariance regression parameters.

In the spirit of GEEs, we formulate a set of joint estimating equations for estimating the heterogeneous mean regression coefficients and correlation regression parameters. Note unlike the classical GEE literature which often treats the working correlation matrix either as known or effectively as a nuisance parameter \citep[e.g.,][]{liangANDzeger1986,quEtAl2000,wangANDcarey2003}, here we consider both mean and correlation components to be equally important in the modeling process. That is, we are interested in performing inference on both the mean regression coefficients and correlation regression parameters, and this requires an asymptotic theory for the joint set of parameters. 

Under a setting where the number of responses can diverge with an increasing number of clusters, we establish estimation consistency and asymptotic normality of the mean regression coefficient estimators and correlation regression parameter estimators, showing that they exhibit differing rates of convergence. We note that the total number of mean regression coefficients can still tend to infinity even if the number of covariates is fixed, since the mean regression coefficients are heterogeneous across responses. Hence, the involvement of both the multivariate mean and correlation regression substantially increases the difficulty in deriving the theoretical properties of the proposed estimators. Indeed, to our knowledge, this article is the first to formally demonstrate consistency and asymptotic normality for the estimator of a joint model specifying heterogeneous mean regression coefficients, where the number of responses is allowed to diverge.

We develop an estimation procedure that iterates between the aforementioned estimating equations to update the mean regression coefficients (along with any dispersion parameters) and correlation regression parameters. For the latter, estimation must be done in a way to obtain an overall valid correlation matrix i.e., positive definite with ones on the diagonals and off-diagonals between -1 and 1. To overcome this challenge, we propose a novel algorithm that adapts the positive definiteness constrained algorithm of \citet{tao2017} to attain a positive definite covariance matrix estimate under the GEE framework first, before standardizing it to produce valid correlation regression parameter estimates.  

\label{page:alternative_method_introduction_start}Simulation studies across a number of response types demonstrate our proposed approach provides  similar performance to several existing GEE-type methods for estimating the mean regression coefficients, but outperforms them in recovering the between-response correlation matrix.\label{page:alternative_method_introduction_end} \label{page:misspecified_similarity_introduction_start}We also investigate the impact of misspecified similarity measures on the proposed estimator, with results demonstrating the robustness of the mean regression coefficient estimators.\label{page:misspecified_similarity_introduction_end} The proposed model is applied to a multivariate abundance dataset in ecology comprising overdispersed counts of 38 Carabidae ground beetle species sampled throughout Scotland, along with information about the environmental covariates of each site and the traits of each species. Results show the beetle species exhibit quite differing relationships with key environmental indicators such as soil pH and land management, while species traits such as beetle total length and breeding season have important effects in driving the correlations between the species. \label{page:sensitivity_analysis_intro_start}These findings are not particularly sensitive to different specifications of trait similarity measures.\label{page:sensitivity_analysis_intro_end}

The rest of this article is organized as follows. Section \ref{sec:model} introduces the joint mean and correlation regression model, a set of associated estimating equations, and asymptotic results for the resulting joint estimator. Section \ref{sec:algorithm} presents details of the iterative estimation procedure, while Sections \ref{sec:numerical_study} and \ref{sec:ground_beetle} discuss results of the simulation study and application to the ground beetle dataset, respectively. Section \ref{sec:conclusion} offers some concluding remarks. All technical conditions, proofs, and additional simulation results are presented in the Appendix and supplementary material.

\section{A Joint Mean and Correlation Regression Model} \label{sec:model}

Let $ \bm{Y}_i = ( Y_{i1},\cdots,Y_{ip} )^\top $ be a $p$-dimensional response vector collected from the $i$-th cluster  for  $i=1,\cdots,n$. \label{page:cluster_terminology_start}In this article, we follow \cite{pourahmadi1999}, \cite{warton2011regularized} and \cite{mullerETAL2013} among others and use $i$ to index the $i$-th ``cluster'', noting related works adopt other terminologies such as ``subject'' in \cite{liangANDzeger1986} and \cite{quEtAl2000}, ``observation'' in \cite{tao2021}, and ``observational unit'' in \cite{hui2023gee} to refer to cluster.\label{page:cluster_terminology_end} \label{page:unbalanced_data_remark_start}Note also that we focus our developments on the balanced data setting i.e., all $n$ clusters have the same number of responses $p$; see Section \ref{sec:conclusion} for a discussion on extensions to the case of unequal number of responses $p_i$ for each of the $i$-th cluster.\label{page:unbalanced_data_remark_end} Let $ \bm{x}_i = (x_{i1},\cdots,x_{id} )^\top  $ denote a set of covariates associated with the $i$-th cluster that we will relate to the mean of the responses, where $x_{i1} = 1$ corresponds to the intercept term for $i=1,\cdots,n$, and $\{\bmW_k = ( w_{j_1 j_2}^{(k)})_{p \times p}: k = 1,\cdots,K\}$ denote a set of $K$ similarity matrices of dimension $p \times p$ that we will link to the correlation matrix of the responses. \label{page:W_observed2_start}The matrices $\bm{W}_k$ may be available directly as part of the data collection process itself, or constructed from auxiliary information vectors $ \bm{z}_j = ( z_{j1},\cdots,z_{jK} )^\top $ associated with the $j$-th response for $j=1,\cdots,p$.\label{page:W_observed2_end} For the latter, each element $w_{j_1 j_2}^{(k)}$ in $ \bmW_k$  measures the similarity between $z_{j_1k}$ and $z_{j_2k}$ for $j_1\neq j_2$. For instance, if $z_{jk}$  is quantitative, then the similarity $w_{j_1 j_2}^{(k)}$ can be defined as $w_{j_1 j_2}^{(k)}=\exp(-|z_{j_1k}-z_{j_2k}|^2)$, whereas if $z_{jk}$ is qualitative we can set $w_{j_1j_2}^{(k)} = 1$ if $z_{j_1 k}$ and $z_{j_2 k}$ have the same categorical level, and $w_{j_1j_2}^{(k)} = 0$ otherwise \citep[see also][]{johnsonANDwichern1992}. For reasons of parameter identifiability, as will be illustrated later on, the diagonals $w_{j j}^{(k)}$ are set to zeros for all $j = 1,\cdots,p$ and $k = 1,\cdots,K$. 
 
The proposed joint mean and correlation regression model is formulated as follows. First, we assume the mean of each response, denoted here as $E ( Y_{ij} ) = \mu_{ij}( \bmbeta_j )$, is related to the covariates as given by
\begin{equation}
g\{\mu_{ij}( \bmbeta_j )\} = \bm{x}_i^\top \bmbeta_j, \text{ for } i=1,\cdots,n \textrm{ and } j=1,\cdots,p, \label{eq:1st_moment_spec}
\end{equation}
where $\bmbeta_j = (\beta_{j1},\cdots,\beta_{jd})^\top $ is a vector of mean regression coefficients that are heterogeneous across responses, and $g(\cdot)$ is a known link function. Next, we specify the second moment of $Y_{ij}$ as $\mathrm{Var}(Y_{ij}) =  h \{ \mu_{ij} ( \bmbeta_j ) ; \phi_j \}$ for $i=1,\cdots,n$ and $j=1,\cdots,p$, where $\phi_j > 0$ are dispersion parameters that are also heterogeneous across $j=1,\cdots,p$, and $h(\cdot)$ is a known function characterizing the mean-variance relationship of the responses. Common examples of link and variance functions include the logit link $g(\mu)= \log \{\mu/(1-\mu)\}$ and the variance function $h(\mu;\phi) = \phi \mu (1-\mu)$ for binary responses, the log link $g(\mu) = \log(\mu)$ and the variance function $h(\mu; \phi)=\phi \mu $ (or $h(\mu; \phi)= \mu + \phi \mu^2$) for (overdispersed) counts, and the identity link $g(\mu) = \mu$ coupled with the constant function $h(\mu;\phi) = \phi$ for continuous responses \citep{fitzmaurice2012applied}. Turning to the correlation regression component of the model, let $\mathrm{Corr}(Y_{ij_1}, Y_{ij_2}) = r_{j_1 j_2}(\bmrho )\; \textrm{for } j_1,j_2 = 1,\cdots,p$, and subsequently define $\bmR ( \bmrho ) =  (  r_{j_1 j_2}(\bmrho ) )_{p \times p} $ as the $p \times p$ (working) correlation matrix of $\bm{Y}_i$ for $i = 1,\cdots,n$. Building upon \citet{tao2017,tao2020,ZOU2021}, we model this correlation matrix as
\begin{equation}
\bmR ( \bmrho ) = \bm{I}_p + \sum_{k=1}^{K} \rho_k \bmW_k,
\label{eq:cor_reg_model}
\end{equation}
where $\bm{I}_p$ is the $ p $-dimensional identity matrix, and $\bmrho=(\rho_1,\cdots,\rho_K)^\top $ is a vector of correlation regression parameters that possess a simple, direct interpretation as quantifying the impact of the similarities on the correlation between responses. For example, in ecology where similarity matrices are constructed from species traits, a higher, positive value of $\rho_k$ implies that, conditional on other traits, two species with more similar values in their $k$-th trait variable are expected to have a stronger positive correlation after accounting for differences in their mean response (which may suggest this trait is important in mediating biotic interactions between species; see the application in Section \ref{sec:ground_beetle}).

Equation \eqref{eq:cor_reg_model} offers a parsimonious yet explicit way to model correlations between responses, as it only involves estimating a vector of $K$ correlation regression parameters, $\bmrho$. This formulation also includes various structured correlation matrices (not driven by the data itself) as special cases, including traditional autoregressive, moving average, compound symmetry, and banded structures \citep[see][]{tao2017}. By denoting $\bmW_0 = \bm{I}_p $ and $\rho_0 = 1$, the correlation regression model can be written as $\bmR ( \bmrho ) =\sum_{k=0}^{K} \rho_k \bmW_k$. As reviewed in Section \ref{sec:intro}, similar ideas of expressing the covariance or correlation matrix as a function of a linear combination of matrices have been considered previously in the literature. Importantly, we consider $\bmrho$ as being equal in importance to $\bmbeta$. This is in contrast to previous studies and variations of GEEs, where the (parameters characterizing the) working correlation have been either treated as a nuisance or used largely to improve the efficiency of inference on the mean model. Therefore, it is imperative to develop asymptotic theory for the joint estimator of all the regression parameters $\bmtheta = ( \bm{\beta}^\top,\bm{\rho}^\top )^\top $, so as to provide a basis for simultaneous inference on the mean and correlation components of the proposed model.
 
\label{page:GEE_framework2_start}Note we model $\bm{R}(\bm{\rho})$ as a correlation instead of a covariance matrix: this is consistent with the requirement of a working correlation matrix in the GEE framework where the response variances are already modeled by the variance function\label{page:GEE_framework2_end}, \label{page:identifiability_start}and also circumvents parameter identifiability problems between the covariance parameters and dispersion parameters when the variance function has a multiplicative relationship with the dispersion parameter e.g., if $h(\mu; \phi) = \phi\mu$ and $\bm{R}(\bm{\rho}) = \rho_0 \bm{I}_p + \sum_{k=1}^{K} \rho_k \bm{W}_k$ is modeled as a covariance matrix, then the covariance parameters $c\rho_0, \cdots, c\rho_K$ and dispersion parameters $\phi_1/c, \cdots, \phi_p /c$ would result in the same covariance structure of $\bm{Y}_i$ vectors for any value of $c > 0$.\label{page:identifiability_end} On the other hand, the estimation of $\bmrho$ is now more challenging compared to previous covariance regression models of \citet{tao2017,tao2020, ZOU2021}. Specifically, since equation \eqref{eq:cor_reg_model} is a regression model for the correlation matrix, $\bmrho$ must be in the parameter space $ \mathscr{P}^+ =\{\bmrho:\bmR ( \bmrho )\textrm{ is a valid correlation matrix}\}$. That is, the diagonals of $\bmR ( \bmrho )$ need to be ones and off-diagonals between -1 and 1, in addition to $\bmR (\bmrho)$ being positive definite. \label{page:specific_example_start}We provide two simple but insightful examples to illustrate this, in the case of a single similarity matrix $\bm{W}_1$. First, if $\bm{W}_1$ is a compound symmetry matrix i.e., $w_{j_1j_2}^{(1)} = c > 0$ for $j_1 \neq j_2$, then $\mathscr{P}^+ = \{\rho_1: \rho_1 \in ( -(p-1)^{-1} c^{-1}, c^{-1}  ) \}$. Second, if $\bm{W}_1$ is a tridiagonal matrix with $w_{j_1j_2}^{(1)} = c \neq 0$ for $| j_1 - j_2| = 1$ and zero otherwise, then the parameter space is $\mathscr{P}^+ = \{\rho_1: \rho_1 \in ( 0.5 |c|^{-1} [\cos\{p\pi/(p+1)\} ]^{-1}, -0.5 |c|^{-1} [\cos\{p\pi/(p+1)\} ]^{-1}  ) \}$. Needless to say, when multiple general similarity matrices are involved the requirement for $\bmrho \in \mathscr{P}^+$ becomes even more complex; this motivates the estimating equations proposed in the following subsection, which will take this requirement into consideration.  \label{page:specific_example_end}

\vspace{-1em}

\subsection{Estimating Equations} \label{subsec:esteqns}
We establish a set of estimating equations that will be used as the basis for estimation and inference with the proposed joint model. To this end, we write the model in a vectorized form as follows. Let $ \bmY = ( \bmY_1^\top ,\cdots, \bmY_n^\top )^\top$ and $\bmbeta = (\bmbeta^{\top}_1,\cdots,\bmbeta^{\top}_p)^\top$ denote the stacked $np$- and $pd$-dimensional vectors of all responses and mean regression coefficients, respectively. Subsequently, we can write $\bmmu_i( \bmbeta ) =(\mu_{i1}( \bmbeta_1 ) ,\cdots,\mu_{ip}( \bmbeta_p ) )^\top$ for $i=1,\cdots,n$ and $\bmmu ( \bmbeta )=(\bmmu_1^\top( \bmbeta ),$ $\cdots,\bmmu_n^\top(\bmbeta))^\top$. Next, let $\bm{A}_i ( \bmbeta )=\mathrm{diag}[h\{ \mu_{i1} ( \bmbeta_1 ) ; \phi_1 \},$ $\cdots,h\{ \mu_{ip}( \bmbeta_p ) ; \phi_p \}]$ denote $p \times p$ diagonal matrices of the variance functions at the $i$-th cluster for $i=1,\cdots,n$, where for ease of notation we have suppressed the dependence on the $\phi_j$'s, and let $\bm{A} ( \bmbeta ) = \mathrm{diag} \{ \bm{A}_1 ( \bmbeta ), \cdots , \bm{A}_n ( \bmbeta )  \}$ be a block diagonal matrix with the $i$-th block being $\bm{A}_i ( \bmbeta )$. By denoting $\mathrm{Cov}( \bmY )$ as the full $np \times np$ covariance matrix of the vector $\bmY$, equations \eqref{eq:1st_moment_spec} -- \eqref{eq:cor_reg_model} can be expressed as
\begin{equation} 
	E(\bmY) = \bmmu ( \bmbeta ), \;
	\mathrm{Cov}( \bmY ) = \bmA^{1/2} ( \bmbeta ) \lbk \bm{I}_{n} \otimes \bmR( \bmrho )\rbk\bmA^{1/2} ( \bmbeta ), 
    \label{eq:moment_spec}
\end{equation}
where $ \otimes $ is the Kronecker product operator and the form of the covariance in equation \eqref{eq:moment_spec} is analogous to that seen in GEEs previously \citep[e.g.,][]{liangANDzeger1986,warton2011regularized}. 

Next, consider the matrix-valued function $\bmSigma ( \bmalpha ) = \alpha_0 \bm{I}_{p} + \sum_{k=1}^{K} \alpha_k \bmW_k,$ where $ \bmalpha = ( \alpha_0,\cdots,  \alpha_K )^\top \in \mathscr{A}^+$ lies in the positive definite parameter space $\mathscr{A}^+ =\{\bmalpha: \bmSigma ( \bmalpha )$ is positive definite$\}$. Recall the diagonals of $\bm{W}_k$ are zeros for $k=1,\cdots, K$. Then we have the following proposition.

\begin{proposition}\label{pn:1} 
If there exists an $\bmalpha\in \mathscr{A}^+$, then $(\alpha_1/\alpha_0, \cdots,   \alpha_K/\alpha_0 )^\top\in \mathscr{P}^+$.
\end{proposition}

\noindent The above result follows directly from the definition of a correlation matrix. Importantly, it implies if we are able to obtain values $\bm{\alpha} \in \mathscr{A}^+$, then a valid (estimated) correlation matrix $\bmR ( \bmrho)$ can be immediately obtained by setting $\rho_1 = \alpha_1/\alpha_0, \cdots, \rho_K = \alpha_K/\alpha_0$, with the resulting values of $\bmrho$ satisfying $\bmrho \in \mathscr{P}^+$. 
Motivated by this, we propose to consider a slight reparameterization of the correlation regression model in equation \eqref{eq:cor_reg_model} as
$
	\bmSigma ( \bmalpha ) = \alpha_0 \bm{I}_{p} + \sum_{k=1}^{K} \alpha_k \bmW_k
	=\bmR ( \bmrho ),
$
with $\alpha_0=1$ and $\alpha_k=\rho_k$  for $ k = 1,\cdots,K$. The covariance equation in (\ref{eq:moment_spec}) can then be rewritten as $\mathrm{Cov} ( \bm{Y} )= \bmAhalf ( \bmbeta ) \{ \bm{I}_n \otimes \bmSigma ( \bmalpha ) \}  \bmAhalf ( \bmbeta )$, and subsequently the parameters we solve for are now $\bmvartheta = (\bmbeta^\top, \bmalpha^\top)^\top$. 

Let $\bm{D} ( \bmbeta ) = \partial \bmmu (\bmbeta )/ \partial \bmbeta^\top \in \mathbb{R}^{np \times pd}$ and $\tildebmSigma (\bmalpha) = \bm{I}_n \otimes \bmSigma(\bmalpha)$. For estimating the mean regression coefficients $\bmbeta$, we consider the estimating equation
\begin{equation}
	\bmpsi_{\bmbeta}\lsk \bmbeta, \bmalpha \rsk = \bm{D}^\top \lsk \bmbeta \rsk \bmANhalf\lsk \bmbeta \rsk \tildebmSigma^{-1} \lsk \bmalpha \rsk \bmANhalf\lsk \bmbeta \rsk \lbk \bmY - \bmmu\lsk \bmbeta \rsk \rbk = \bm{0}_{pd},
	\label{eq:psi_beta}
\end{equation}
where $ \bm{0}_{pd}$ is a $pd$-dimensional vector of zeros. \label{page:robustness_beta_start}Equation \eqref{eq:psi_beta} for the mean regression coefficients is the same as in GEEs, meaning we have $\rmE \{\bmpsi_{\bmbeta}\lsk \bmbeta, \bmalpha \rsk \}= \bm{0}_{pd}$ when the estimating equation is evaluated at the true values of the mean regression coefficients regardless of the specification of $\tildebmSigma^{-1}(\bmalpha)$.\label{page:robustness_beta_end} Next, let $ \bmepsilon_i ( \bmbeta ) = \bmA^{-1/2}_{i} ( \bmbeta ) \{ \bmY_i - \bmmu_i (\bmbeta) \}$ for $i=1,\cdots,n$, $\bmepsilon( \bmbeta ) = (\bmepsilon_1^\top( \bmbeta ) ,\cdots ,  \bmepsilon_n^\top ( \bmbeta ))^\top$, and $ \tilde{\bmW}_k = \bm{I}_{n} \otimes \bmW_k$ for $ k = 0,\cdots,K $. Then for the reparameterized correlation regression parameters $\bmalpha$, we consider solving the estimating equation
\begin{equation}
	\bmpsi_{\bmalpha} \lsk \bmbeta, \bmalpha \rsk = \lsk \bmepsilon^\top \lsk \bmbeta \rsk \tilde{\bmW}_k \bmepsilon\lsk \bmbeta \rsk   \rsk_{(K+1) \times 1} - \lsk \tr ( \tilde{\bmW}_{k_1} \tilde{\bmW}_{k_2} ) \rsk _{(K+1) \times (K+1)}\bmalpha = \bm{0}_{K+1},
	\label{eq:psi_alpha}
\end{equation}
where $ ( \tr ( \tilde{\bmW}_{k_1} \tilde{\bmW}_{k_2} ) )_{(K+1) \times (K+1)}  $ denotes a $ (K+1) \times (K+1) $ matrix whose $ (k_1+1,k_2+1) $-th element is given by $ \tr ( \tilde{\bmW}_{k_1} \tilde{\bmW}_{k_2} ) $ for $ k_1,k_2 = 0,\cdots,K $, and $( \bmepsilon^\top ( \bmbeta ) \tilde{\bmW}_k \bmepsilon( \bmbeta )   )_{(K+1) \times 1}$ is a $(K+1)$-dimensional vector whose $ (k+1) $-th element is $ \bmepsilon^\top ( \bmbeta ) \tilde{\bmW}_k \bmepsilon( \bmbeta ) $ for $ k= 0,\cdots,K $. Equation \eqref{eq:psi_alpha} is related to the first order condition of the least squares optimization problem considered by \cite{tao2017,tao2020}, who estimated covariance regression parameters by minimizing the least squares loss function $Q(\bmbeta,\bmalpha)=\sum_{i=1}^{n} \left\| \bmepsilon_i ( \bmbeta ) \bmepsilon_i^\top ( \bmbeta ) - \bmSigma ( \bmalpha )  \right\|_F^2$ with respect to $\bmalpha$ when $\bmbeta$ is given, where $ \| \bm{H} \|_F = \{ \tr ( \bm{H}^\top \bm{H} ) \}^{1/2}$ is the Frobenius norm for a generic matrix $ \bm{H} $. Specifically, the solution of equation \eqref{eq:psi_alpha} satisfies the first order condition of the optimization problem $\min_{\bmalpha} Q(\bmbeta,\bmalpha)$ since $-2\bmpsi_{\bmalpha}(\bmbeta,\bmalpha) = \partial Q(\bmbeta,\bmalpha)/\partial \bmalpha$. It can also be verified that $\rmE \{ \bmpsi_{\bmalpha} ( \bmbeta, \bmalpha )\} = \bm{0}_{K+1} $ where $\bmpsi_{\bmalpha}$ is evaluated at the true values of the mean regression coefficients and the reparameterized correlation regression parameters. Other forms of estimating equation could be considered for the reparameterized correlation regression parameters e.g. see Section 3.3.1 of \cite{lipsitzANDfitmaurice2008}, although we focus our developments on equation \eqref{eq:psi_alpha} in this article.

By solving the joint estimating equation $\bmpsi(\bmvartheta)=(\bmpsi_{\bmbeta}^\top ( \bmbeta, \bmalpha ), \bmpsi_{\bmalpha}^\top ( \bmbeta, \bmalpha ))^\top = \bm{0}_{pd+K+1}$, we obtain the estimator $\hat{\bmvartheta} = ( \hat{\bmbeta}^\top, \hat{\bmalpha}^\top )^\top $. Afterward, following Proposition \ref{pn:1} we obtain the joint estimator of the mean regression coefficients and the correlation regression parameters as $\hat{\bmtheta} = ( \hat{\bmbeta}^\top, \hat{\bmrho}^\top )^\top $, where $\hat\rho_1 = \hat\alpha_1 / \hat\alpha_0,\cdots, \\ \hat\rho_K = \hat\alpha_K/\hat\alpha_0$.

\subsection{Asymptotic Theory} \label{sec:asym}
We first study the asymptotic properties of $\hat{\bmvartheta} = ( \hat{\bmbeta}^\top, \hat{\bmalpha}^\top )^\top$ as the number of clusters $n \to \infty$ and the number of responses $p$ can grow with increasing $n$, with the dispersion parameters assumed to be known. We emphasize that although both $d$ and $K$ are assumed to be fixed, $\hat{\bmvartheta}$ has a growing dimension as $p \to \infty $ since the dimension of $\bmbeta$ grows with order $p$. To facilitate developments of the asymptotic theory, we introduce a matrix
$
\bmXi^{(\mS)} = \mathrm{diag}(\bm{T}^{(\mS)} \otimes \bm{I}_d, \bm{I}_{K+1}),
$
where $\mS=\{s_1,\cdots,s_q\}$ generically denotes a subset of $ \{1,\cdots,p\}$ with finite size $q$, $\bm{T}^{(\mS)} = (\bm{t}_{s_1},\cdots, \bm{t}_{s_q})^\top$ with $\bm{t}_{s_j}$ being the $s_j$-th column of the identity matrix $\bm{I}_{p}$ for $ j = 1,\cdots,q $, and $ \bm{0}_{k_1\times k_2}$ denotes a $k_1\times k_2$ matrix of zeros. It follows that $\bmXi^{(\mS)} \hat{\bmvartheta} = ( \hat{\bmbeta}_{s_1}^{\top},\cdots, \hat{\bmbeta}_{s_q}^{\top}, \hat{\bmalpha}^{\top})^\top$ is a finite dimensional sub-vector of $\hat{\bmvartheta}$. 

Let $\bmvarthetao = (\bmbeta^{(0)\top},\bmalpha^{(0)\top})^\top$ denote the true parameter value for $\bmvartheta$, where $\bmSigma(\bmalphao)$ is positive definite. In the following theorem, the estimators $(\hat{\bmbeta}_{s_1}^{\top},\cdots, \\ \hat{\bmbeta}_{s_q}^{\top})^\top$ and $\hat{\bmalpha}$ have differing rates of convergence towards their respective true values. To accommodate this, we define $ \bmmA = \mathrm{diag}(\sqrt{n} \bm{I}_{qd}, \sqrt{np} \bm{I}_{K+1} ) $ and proceed to derive the asymptotic distribution of the quantity $ \bmmA \bmXi^{(\mS)} ( \hat{\bmvartheta} - \bmvarthetao )$. Let $ \tbmmA = \mathrm{diag}(\sqrt{n} \bm{I}_{pd}, \sqrt{np} \bm{I}_{K+1} ) $ and define the matrices
$
\bm{B} = \tbmmA^{-1} E \{ \partial \bmpsi(\bmvarthetao)/\partial \bmvartheta^\top \\ \} \tbmmA^{-1}$, $  \bm{U} = \tbmmA^{-1} \mathrm{Cov}\{\bmpsi(\bmvarthetao)\} \tbmmA^{-1}$ and $ 
\bm{\Omega}(\bmXi^{(\mS)}) = \bmXi^{(\mS)} \bm{B}^{-1}  \bm{U}\bm{B}^{-1 \top} \bmXi^{(\mS)\top},
$
where closed-form expressions for both $E \{ \partial \bmpsi(\bmvarthetao)/\partial \bmvartheta^\top\}$ and $\mathrm{Cov}\{\bmpsi(\bmvarthetao)\}$ are provided in Lemma \ref{lemma:closed_form} of supplementary material \ref{appendix:lemmas} . We then have the following result for the estimator $\hat{\bmvartheta} = (\hat{\bmbeta}^\top, \hat{\bmalpha}^\top)^\top$.
 
\begin{theorem}\label{theorem:mean_cov}
Under Conditions \ref{condition:varepsilon} -- \ref{condition:B_and_Omega} in the Appendix, it follows that
$
	\bm{\Omega}^{-1/2}( \bmXi^{(\mS)} ) \bmmA \\\bmXi^{(\mS)}  ( \hat{\bmvartheta} - \bmvarthetao ) \stackrel{d}{\longrightarrow} N (\bm{0}_{qd+K+1} , \bm{I}_{qd+K+1}),
$
as $n\to \infty$ and $p = o(n^{1/2})$.
\end{theorem}
 
The selection of the subset $\mS=\{s_1,\cdots,s_q\}\subset \{1,\cdots,p\}$ and its size $q$ is arbitrary in Theorem \ref{theorem:mean_cov}, so long as $q$ is finite.  Accordingly, for any given $\mS$, Theorem \ref{theorem:mean_cov} provides the joint limiting distribution for $( \hat{\bmbeta}_{s_1}^{\top},\cdots, \hat{\bmbeta}_{s_q}^{\top}, \hat{\bmalpha}^{\top})^\top$. 
Furthermore, the theorem implies $( \hat{\bmbeta}_{s_1}^{\top},\cdots, \hat{\bmbeta}_{s_q}^{\top})^\top$ and $\hat{\bmalpha}$ are $ \sqrt{n} $- and $ \sqrt{np} $- consistent, respectively. This is not an overly surprising result: since the mean regression coefficients are heterogeneous across responses, then only $n$ observations $\{Y_{ij}: i = 1,\cdots,n\}$ contribute information to the estimation of each $\bmbeta_{j}$. By contrast, $\bmalpha$ are parameters common across all responses, and so its estimation leverages information across both $i=1,\cdots,n$ and $j=1,\cdots,p$. 

Based on the finding in Theorem \ref{theorem:mean_cov}, we have the following result for $\hat{\bmalpha}$.

\begin{corollary}
Under Conditions \ref{condition:varepsilon} -- \ref{condition:B_and_Omega} in the Appendix, it follows that
$
	\mathrm{P} (\hat{\bmalpha} \in  \mathscr{A}^+) \to 1,
$
as $n\to \infty$ and $p = o(n^{1/2})$.
\label{cor:positive_definite}
\end{corollary}

Corollary \ref{cor:positive_definite} provides a theoretical guarantee that the estimated reparameterized correlation regression parameters $\hat{\bmalpha}$ from the joint estimator $\hat{\bmvartheta}$ will fall into the required parameter space $\mathscr{A}^+$ with probability tending to one. This allows us to apply Proposition \ref{pn:1} to obtain the estimated correlation regression parameters $\hat{\bmrho} \in \mathscr{P}^+$ via the transformation $\hat\rho_1 = \hat\alpha_1/\hat\alpha_0,\cdots,\hat\rho_K=\hat\alpha_K/\hat\alpha_0$.  

We now proceed to the estimated parameter vector of interest, namely $\hat{\bmtheta} = ( \hat{\bmbeta}^{\top}, \hat{\bmrho}^{\top} )^\top $. Based on the transformation used to obtain $\hat{\bmrho}$ from $\hat{\bmalpha}$, we can apply techniques similar to the multivariate delta method with the vector-valued function $\bm{f}(\bmvartheta) =({\bmbeta}^{\top}, \alpha_1/\alpha_0,\cdots,$ $\alpha_K/\alpha_0 )^\top $ as follows. Let $\bmthetao = (\bmbeta^{(0)\top},\bmrho^{(0)\top})^\top$ denote the true parameter value for $\bmtheta$, where $\bm{R}(\bmrho^{(0)})$ is a valid correlation matrix, and define $ \barbmXi^{(\mS)} = \mathrm{diag}(\bm{T}^{(\mS)} \otimes \bm{I}_d, \bm{I}_{K}) $ such that $\barbmXi^{(\mS)} \hat{\bmtheta} = ( \hat{\bmbeta}_{s_1}^{\top},\cdots, \hat{\bmbeta}_{s_q}^{\top}, \hat{\bmrho}^{\top})^\top$ is a finite dimensional sub-vector of $\hat{\bmtheta}$. Furthermore, let $\barmathbcalA = \mathrm{diag}(\sqrt{n} \bm{I}_{qd},  \sqrt{np} \bm{I}_{K})$ and 
$
\bar{\bm{\Omega}}(\barbmXi^{(\mS)}) = \barbmXi^{(\mS)} \{ \partial \bm{f}(\bmvarthetao)/\partial \bmvartheta^\top\} \bm{B}^{-1} \bm{U} \bm{B}^{-1 \top} \{ \partial \bm{f}(\bmvarthetao)/\partial \bmvartheta^\top \}^\top \barbmXi^{(\mS)\top}. 
$
We now state our main result for the joint estimator $\hat{\bmtheta} = ( \hat{\bmbeta}^\top, \hat{\bmrho}^\top )^\top $.

\begin{theorem}	\label{theorem:mean_cor}
	Under Conditions \ref{condition:varepsilon} -- \ref{condition:B_and_Omega} in the Appendix, it follows that
	$
		\bar{\bm{\Omega}}^{-1/2}(\barbmXi^{(\mS)}) \barmathbcalA \\ \barbmXi^{(\mS)} ( \hat{\bmtheta} - \bmthetao ) \stackrel{d}{\longrightarrow} N (\bm{0}_{qd+K} , \bm{I}_{qd+K}), 
        $
as $n\to \infty$ and $p = o(n^{1/2})$.
\end{theorem} 

Theorem \ref{theorem:mean_cor} is valid under both cases of fixed $p$ and diverging $p$ as long as $p / \sqrt{n} \to 0$. The theorem shows any finite dimensional sub-vector of the estimator $\hat{\bmtheta}$ is consistent and asymptotically normally distributed, and forms the basis by which we can perform inference on the mean regression coefficients and correlation regression parameters of the proposed model. Moreover, Theorem \ref{theorem:mean_cor} (again) implies $( \hat{\bmbeta}_{s_1}^{\top},\cdots, \hat{\bmbeta}_{s_q}^{\top})^\top$ and $\hat{\bmrho}$ are $ \sqrt{n} $- and $ \sqrt{np} $- consistent, respectively, and that the asymptotic covariance matrix of $(\hat{\bmbeta}_{s_1}^{\top},\cdots, \hat{\bmbeta}_{s_q}^{\top}, \hat{\bmrho}^\top )^\top$ is $\bm{G}^{(\mS)} =  \barmathbcalA^{-1} \bar{\bm{\Omega}}(\barbmXi^{(\mS)}) \barmathbcalA^{-1}$. In practice, to apply the above theorem we need a consistent estimator of $\bm{G}^{(\mS)}$ whose expression (see Lemma \ref{lemma:closed_form} in supplementary material \ref{appendix:lemmas}) turns out to be dependent on $\bmvarthetao$, the third-order moment $\mu^{(3)}$, and the fourth-order moment $\mu^{(4)}$ defined in Condition \ref{condition:varepsilon}. From this, we can verify that a consistent estimator can be obtained by replacing $\bmvarthetao$, $\mu^{(3)}$ and $\mu^{(4)}$ in $\bm{G}^{(\mS)}$ by $\hat\bmvartheta$, the third and fourth-order empirical moments, respectively; see also supplementary material \ref{sec:inference} for details on this.

\section{Estimation Procedure} \label{sec:algorithm}

We develop an iterative estimation procedure to compute the joint estimator of $\bmbeta$ and $\bmrho$ based on the two sets of estimating equations in Section \ref{subsec:esteqns}, while allowing for the estimation of (potentially unknown) dispersion parameters and ensuring the estimators of $\bmalpha$ (and $\bmrho$) are in the required parameter space. 

The estimation procedure consists of iterating between the following two steps. Given $\bmalpha$, we employ a Fisher scoring method, with a tuning parameter $\gamma > 0$ to adjust the step size if appropriate, to solve $ \bmpsi_{\bmbeta} ( \bmbeta, \bmalpha ) = \bm{0}_{pd} $. This leads to an update of the form $\bmbeta \Leftarrow \bmbeta - \gamma \bm{J}^{-1}(\bmbeta, \bmalpha) \bmpsi_{\bmbeta} (\bmbeta, \bmalpha)$ where $\bm{J}( \bmbeta ,\bmalpha) = E\{\partial \bmpsi_{\bmbeta} ( \bmbeta ,\bmalpha)/\partial \bmbeta^\top\} = -\bm{D}^\top(\bmbeta) \bmANhalf(\bmbeta)  \tildebmSigma^{-1}(\bmalpha) \bmANhalf(\bmbeta) \bm{D}(\bmbeta)$. Next, given $\bmbeta$, directly solving $ \bmpsi_{\bmalpha} ( \bmbeta, \bmalpha ) = \bm{0}_{K+1}$ leads to the closed-form solution 
$\bmalpha = ( \tr(\tilde{\bmW}_{k_1} \tilde{\bmW}_{k_2}) )_{(K+1) \times (K+1)}^{-1} ((\bmepsilon^\top({\bmbeta}) \tilde{\bmW}_k  \bmepsilon({\bmbeta}) ) _{(K+1) \times 1}. $
This is equivalent to the solution of the unconstrained least squares optimization problem $\min_{\bmalpha}Q(\bmbeta,\bmalpha)$ defined below equation \eqref{eq:psi_alpha}. However, while simple to compute, this unconstrained estimator is not guaranteed to be in the required parameter space $\mathscr{A}^+$ in finite samples as Corollary \ref{cor:positive_definite} is an asymptotic result. As such, given $\bmbeta$, we instead consider solving the constrained least squares optimization problem
\begin{equation}\label{eq:alpha-step}
\min_{\bmalpha\in\mathscr{A}^+} Q(\bmbeta,\bmalpha),
\end{equation}
which ensures the resulting estimator of $\bmalpha$ is in the parameter space $\mathscr{A}^+$. This constrained optimization can be solved, for instance, using the alternating direction method of multipliers (ADMM) algorithm, \label{page:ADMM_explanation_start}which essentially converts \eqref{eq:alpha-step} into an equivalent optimization problem $\min_{\bmalpha, \bm{\Delta}} \{ Q(\bmbeta,\alpha): \bmSigma(\bmalpha) = \bm{\Delta}, \bm{\Delta} - \nu \bm{I}_p \textrm{ is positive definite} \}$ by introducing a $p \times p$ augmented parameter matrix $\bm{\Delta}$, where $\nu$ is an arbitrarily small positive constant to guarantee positive definiteness. The solution of \eqref{eq:alpha-step} is then obtained by minimizing the augmented Lagrangian function of the latter optimization problem through an iterative way; see supplementary material \ref{sec:admm} for details of the iterative steps in the ADMM algorithm, as well as \cite{xueETAL2012} and \cite{tao2017} for more general discussion of the algorithm.\label{page:ADMM_explanation_end} If the unconstrained estimator is in $\mathscr{A}^+$, the ADMM algorithm will end up producing the same estimator as the unconstrained estimator. Therefore, in practice it is computationally efficient to first compute the unconstrained estimator and check if it is in $\mathscr{A}^+$. Only when this is not satisfied do we proceed to solve \eqref{eq:alpha-step}. Finally, if required we can estimate each element $\phi_j$ of the dispersion parameter vector $\bmphi = (\phi_1,\cdots,  \phi_p)^\top$ by solving the estimating equation based on the moment condition
$
E[   \{Y_{ij}-\mu_{ij}(\bmbeta_j)\}^2/h \{ \mu_{ij}(\bmbeta_j);\phi_j \} ] = 1,
$
where the means $\mu_{ij}$ are evaluated at the true mean regression coefficients.  

\label{page:algorithm_explanation_start}A formal algorithm detailing the estimation procedure is provided in supplementary material \ref{sec:admm}. To summarize, it begins by estimating $\bmbeta$ using the above Fisher scoring method given some initial values of $\bmbeta$ and $\bmalpha$, before computing the unconstrained estimator of $\bmalpha$ by solving $\min_{\bmalpha} Q(\bmbeta,\bmalpha)$ given the updated $\bmbeta$. If the unconstrained estimator is not in the parameter space $\mathscr{A}^+$, then the ADMM algorithm is used to solve the constrained optimization problem \eqref{eq:alpha-step}. Afterward, the estimated $\bmalpha$ is transformed into $\bmrho$ by following Proposition \ref{pn:1}, and the iterative updates of $\bmbeta$ and $\bmrho$ proceed until convergence.\label{page:algorithm_explanation_end}

\section{Simulation Study} \label{sec:numerical_study}

To assess the finite sample performance of the proposed estimator, we conducted a simulation study by generating correlated multi-response data from Bernoulli, Poisson, negative binomial, and Gaussian distributions, and considering different sample sizes $ n \in \{50,100,200,400\} $ and number of responses $ p \in \{10,25,50\} $ with $d = 4$ covariates and $K = 5$ similarity matrices.  A total of 1000 datasets are simulated for each combination of responses distribution, $n$ and $p$. Full details of the simulation setup are provided in supplementary material \ref{subsec:simulation_setting}. To assess performance, we computed the mean square error (MSE) of the mean regression coefficient estimators $ \mathrm{MSE}(\bmbeta) =  \sum_{j=1}^{p} \sum_{l=1}^{d} (\hat{\beta}_{jl} - \betao_{jl})^2/pd $, and the correlation regression parameter estimators $ \mathrm{MSE}(\bmrho) =  \sum_{k=1}^{K} (\hat{\rho}_{k} - \rhoo_k)^2/K $, noting the MSE was scaled by number of associated parameters, where $\hat{\beta}_{jl}$ and $\hat{\rho}_{k}$ denote the corresponding elements of $\hat{\bmbeta}$ and $\hat{\bmrho}$, respectively, from the estimation procedure in Section \ref{sec:algorithm}, while $\betao_{jl}$ and $\rhoo_{jl}$ denote the true values.

\begin{figure}[tb]
    \begin{center}
	\includegraphics[width=0.75\textwidth]{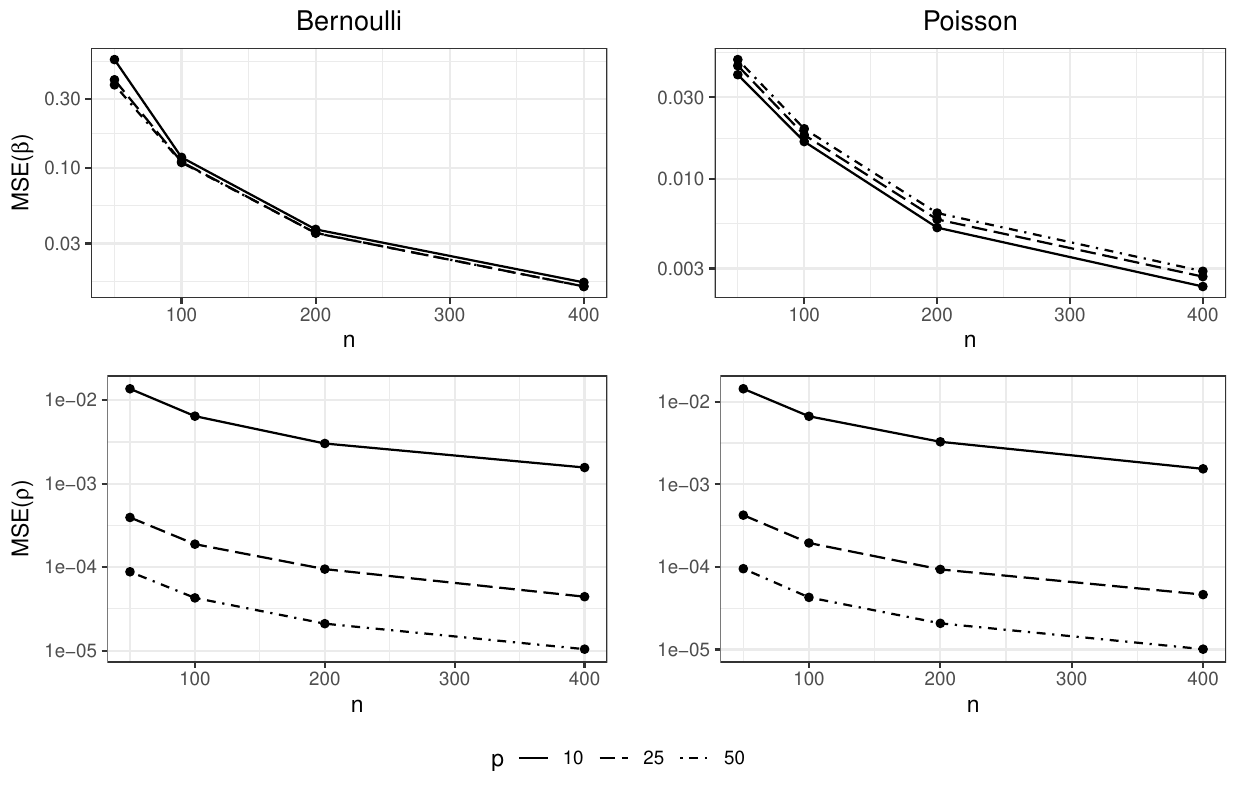}\par
    \end{center} \vspace{-2em}
        \caption{Averaged MSE of the mean regression coefficient estimators (top row) and correlation regression parameter estimators (bottom row) for Bernoulli (left column) and Poisson (right column) responses.} \vspace{-1em} 
	\label{figure:simulation}
\end{figure}

For brevity, we present results for the case of Bernoulli and Poisson responses here; the results for negative binomial and Gaussian responses are provided in supplementary material \ref{sec:additional_simulation} and present very similar conclusions. Figure \ref{figure:simulation} shows the averaged MSE (over 1000 replications) for both $\bmbeta$ and $\bmrho$ decreased when $ n $ increased. Furthermore, the averaged MSE of $\bmrho$ decreased when $ p $ increased, but the averaged MSE of $\bmbeta$ was relatively unaffected by $ p $. This is not surprising given $\bmbeta_j$'s were heterogeneous across responses in the proposed model. By contrast, additional responses provided more information about the correlation structure, leading to improved estimation of $\bmrho$. The differing behaviour of the averaged MSE for $\bmbeta$ and $\bmrho$ supports our theoretical findings in Theorem \ref{theorem:mean_cor} that the mean regression coefficient estimators and correlation regression parameter estimators have different convergence rates i.e., $ \sqrt{n} $ and $ \sqrt{np} $, respectively. Finally, one interesting finding from our simulation study is that across all 1000 replications for various settings of responses distribution, $n$ and $p$, the unconstrained estimator discussed in Section \ref{sec:algorithm} was always in the parameter space $\mathscr{A}^+$. This aligns with Corollary \ref{cor:positive_definite} and reinforces the idea of first computing the unconstrained estimator and checking whether it produces a positive definite covariance matrix. \label{page:algorithm_runtime1_start}Turning to computational efficiency, Table \ref{table:algorithm_runtime} in supplementary material \ref{sec:additional_simulation} provides details of the mean runtime for our proposed estimation algorithm, from which we see convergence was exceptionally quick (taking less than three seconds on average) using an AMD Ryzen 7 CPU @ 3.60 GHz machine.\label{page:algorithm_runtime1_end}

Next, we studied the inferential performance of the joint estimator. Recall the definition of $\bm{G}^{(\mS)}$ given below Theorem \ref{theorem:mean_cor}, which we further represent as a block matrix below,
\begin{equation*}
\bm{G}^{(\mS)}  = \begin{pmatrix}
	{\bm{V}}_{\bmbeta_{\mS}} & {\bm{V}}_{\bmbeta_{\mS} \bmrho} \\
	{\bm{V}}_{\bmrho \bmbeta_{\mS} } & {\bm{V}}_{\bmrho}
\end{pmatrix}.
\end{equation*}
Denoting $\hat{\bmbeta}_{\mS} = (\hat{\bmbeta}_{s_1}^{\top},\cdots,\hat{\bmbeta}_{s_q}^{\top})^\top$ and $\bmbetao_{\mS} = (\bmbeta_{s_1}^{(0)\top},\cdots,\bmbeta_{s_q}^{(0)\top})^\top$, Theorem \ref{theorem:mean_cor} implies the quantities $ (\hat{\bmbeta}_{\mS}-\bmbetao_{\mS})^\top {\bm{V}}_{\bmbeta_{\mS}}^{-1}(\hat{\bmbeta}_{\mS} -\bmbetao_{\mS})  $ and $(\hat{\bmrho} -\bmrhoo)^\top {\bm{V}}_{\bmrho}^{-1}(\hat{\bmrho} -\bmrhoo) $ asymptotically follow $ \chi^2 $ distributions with degrees of freedom $ qd $ and $ K $, respectively, as  $n\to \infty$ and $p = o(n^{1/2})$. Based on this, we investigated the empirical coverage probability of 95\% confidence regions for $ \bmbetao_{\mS} $ and $ \bmrhoo $ by considering the mean regression coefficients for the first five responses i.e., $ \mS = \{1,\cdots,5\} $, and  computing the proportion of simulated datasets in which $ (\hat{\bmbeta}_{\mS} -\bmbetao_{\mS})^\top \hat{\bm{V}}_{\bmbeta_{\mS}}^{-1}(\hat{\bmbeta}_{\mS} -\bmbetao_{\mS}) \leq \chi^2_{qd,0.95} $ and $(\hat{\bmrho} -\bmrhoo)^\top \hat{\bm{V}}_{\bmrho}^{-1}(\hat{\bmrho} -\bmrhoo) \leq \chi^2_{K,0.95}$. Here, $ \chi^2_{l,0.95} $ denotes the 95\% quantile of the $ \chi^2 $ distribution with degrees of freedom $ l $, and $ \hat{\bm{V}}_{\bmbeta_{\mS} }$ and $ \hat{\bm{V}}_{\bmrho} $ are the estimators of $ \bm{V}_{\bmbeta_{\mS}}  $ and $ \bm{V}_{\bmrho} $, respectively, obtained from the corresponding blocks of the consistent estimator $\hat{\bm{G}}^{(\mS)}$ of $\bm{G}^{(\mS)}$; see supplementary material  \ref{sec:inference} for further details on computing $\hat{\bm{G}}^{(\mS)}$. Table \ref{table:chisq_coverage} demonstrates that the resulting coverage of 95\% confidence regions for $ \bmbetao_{\mS} $ tended to the nominal level of 95\% as $ n $ increased, but was relatively unaffected by $p$. There was slight overcoverage for the case of Bernoulli responses. The coverage of 95\% confidence regions for $\bmrhoo$ tended to 95\% as $ n $ and/or $ p $ increased. In supplementary material \ref{sec:additional_simulation}, we present additional results for the empirical coverage probability of 95\% confidence intervals for the mean regression coefficients and correlation regression parameters individually, and obtain similar conclusions.

\begin{table}[t!]                
\caption{Empirical coverage probability for 95\% confidence regions of $ \bmbetao_{\mS} $ and $ \bmrhoo $ for Bernoulli and Poisson responses, where $\mS =\{1,\cdots,5\}$.} \vspace{-0.5em}    \centering \medskip
	\resizebox{0.73\linewidth}{!}{\begin{tabular}{|l l r r r l r r r|}                 
 \hline
		\multicolumn{2}{|r}{} & \multicolumn{3}{c}{Bernoulli} & & \multicolumn{3}{c|}{Poisson}  \\  
  \hline
  & &$ p = 10 $ 	& $ p = 25 $	& $ p = 50 $ & & $ p = 10 $ 	& $ p = 25 $	& $ p = 50 $ \\
		\hline  
		\multirow{4}{*}{$ \bmbetao_{\mS} $}          
  	&$n=50$ & 0.957 & 0.955 & 0.958 & & 0.851 & 0.856 & 0.857 \\ 
		&$n=100$ & 0.966 & 0.968 & 0.967 & & 0.927 & 0.929 & 0.928 \\ 
		&$n=200$ & 0.969 & 0.969 & 0.970 & & 0.933 & 0.936 & 0.934 \\ 
		&$n=400$ & 0.968 & 0.968 & 0.968 & & 0.941 & 0.943 & 0.943 \\  
		\hline     
		\multirow{4}{*}{$ \bmrhoo $}                
		&$n=50$ & 0.852 & 0.874 & 0.886 & & 0.857 & 0.858 & 0.880 \\ 
		&$n=100$ & 0.901 & 0.920 & 0.918 & & 0.917 & 0.909 & 0.912 \\ 
		&$n=200$ & 0.932 & 0.930 & 0.933 & & 0.924 & 0.925 & 0.940 \\ 
		&$n=400$ & 0.930 & 0.945 & 0.933 & & 0.924 & 0.941 & 0.950 \\  
		\hline
	\end{tabular}}   \vspace{-1em}                                                                                                          
	\label{table:chisq_coverage}                                                                                                                                       
\end{table}

\label{page:alternative_method_main_start}
In supplementary material \ref{subsec:compare_other_method}, we present results comparing our proposed estimator to existing GEE methods assuming either an independence working correlation matrix, or an unstructured working correlation matrix. Overall, the proposed estimator was shown to have comparable performance to these existing methods in terms of estimating $\bmbeta$
(this is not surprising given the robustness of the mean regression coefficient estimators to the specification of working correlation structure), 
but clearly outperformed GEE with unstructured working correlation matrix in recovering the true correlation matrix given it leverages the additional information available from the similarity measures. \label{page:alternative_method_main_end}\label{page:misspecified_similarity_main_start}Finally, we carried out numerical studies in supplementary material \ref{subsec:misspecified_similarity} to investigate the impact of misspecifying similarity measures on the estimation performance of the proposed estimator. Results show the estimation of $\bmbeta$ was again largely unaffected by this misspecification, while the estimation performance for $\bmrho$ and hence recovery of the true correlation matrix exhibited a small deterioration in performance.\label{page:misspecified_similarity_main_end}

\section{Application to Scotland Carabidae Ground Beetle Dataset} \label{sec:ground_beetle}
We applied the proposed joint mean and correlation regression model to a multivariate abundance dataset from ecology comprising overdispersed counts of Carabidae ground beetle species in Scotland. The data was sampled from a total of $n=87$ sites spread across nine main areas in Scotland using pitfall traps \citep{riberaETAL2001}, with the aim of the study being to jointly quantify the effects of environmental processes and trait mediation on the ground beetle assemblages; see \cite{riberaETAL2001} for more details of the study. For illustrative purposes, we considered a subset of $p=38$ carabid ground beetle species that were detected in at least 15 sites. Along with species abundance observations, we considered three environmental covariates in soil pH, elevation above sea level, and land management intensity score \citep{downieETAL1999}, all of which were centered and scaled to have zero mean and unit variance prior to analysis. Together with the intercept, this leads to $d=4$ predictors in $\bm{x}_i$. \label{page:K_equal_5_start}For each species, we also have records for $K=5$ trait predictors, with one being quantitative (total length) and four being qualitative (color of the legs, wing development, overwintering and breeding season). We converted each of the trait predictors into a similarity matrix based on the procedure discussed above equation \eqref{eq:1st_moment_spec}, depending on whether it is a quantitative or qualitative trait.\label{page:K_equal_5_end} 

From an exploratory analysis (see Figure \ref{figure:log_log_beetle} in supplementary material \ref{sec:supp_ground_beetle}), we observed evidence of overdispersion and a quadratic mean-variance relationship for the $p=38$ species. Along with the fact that different carabid beetle species are known to exhibit diverse responses to the environment, we thus proceeded to fit the joint mean and correlation regression model assuming a log link function for the mean, and a quadratic variance function $h(\mu;\phi) = \mu + \phi \mu^2 $ where the species-specific overdispersion parameters $\phi_j > 0$ were estimated using the estimation procedure in Section \ref{sec:algorithm}. Based on Theorem \ref{theorem:mean_cor}, we also constructed 95\% confidence intervals for each of the mean regression coefficients and correlation regression parameters; see supplementary material \ref{sec:inference} for further details on this construction.

\begin{table}[t!]
\caption{Point estimates and 95\% confidence intervals (in parentheses) for mean regression coefficients of the first ten species, and correlation regression parameters for all five trait variables, based on analysis of the Carabidae ground beetle dataset. Estimates whose confidence interval excludes zero are bolded.} \centering \medskip
	\resizebox{0.8\linewidth}{!}{\begin{tabular}{|c c c c c|}     
		\hline
		& &Estimation of $\bmbeta_j$ & & \\
		\hline                
		Species & Intercept & Soil pH & Elevation & Land Management \\ 
		\hline
		\multirow{2}{*}{\emph{A.Muelleri}} & 	\textbf{2.444} & 	\textbf{1.382} & 	\textbf{-0.730} & -0.022 \\ 
		& 	\textbf{(2.015, 2.872)} & 	\textbf{(0.631, 2.132)} & 	\textbf{(-1.396, -0.064)} & (-0.725, 0.682) \\ 
		\multirow{2}{*}{\emph{A.Apricaria}} & -0.697 & -0.151 & 	\textbf{-1.649} & 	\textbf{1.615} \\ 
		& (-1.528, 0.135) & (-1.186, 0.883) & 	\textbf{(-2.904, -0.393)} & 	\textbf{(0.692, 2.538)} \\ 
		\multirow{2}{*}{\emph{A.Bifrons}} & 	\textbf{-1.349} & 	\textbf{1.415} & 	\textbf{-2.145} & 0.314 \\ 
		& 	\textbf{(-2.462, -0.236)} & 	\textbf{(0.134, 2.696)} & 	\textbf{(-3.720, -0.570)} & (-0.775, 1.404) \\ 
		\multirow{2}{*}{\emph{A.Communis}} & 	\textbf{0.604} & -0.080 & -0.439 & 	\textbf{-0.970} \\ 
		& 	\textbf{(0.145, 1.062)} & (-0.876, 0.715) & (-1.058, 0.181) & 	\textbf{(-1.741, -0.198)} \\ 
		\multirow{2}{*}{\emph{A.Familiaris}} & 0.221 & 0.454 & -0.371 & 0.474 \\ 
		& (-0.382, 0.823) & (-0.593, 1.500) & (-1.287, 0.544) & (-0.509, 1.458) \\ 
		\multirow{2}{*}{\emph{A.Lunicollis}} & -0.016 & -0.714 & 	\textbf{-0.841} & -0.542 \\ 
		& (-0.533, 0.502) & (-1.644, 0.217) & 	\textbf{(-1.600, -0.082)} & (-1.401, 0.317) \\ 
		\multirow{2}{*}{\emph{A.Plebeja}} & 	\textbf{3.074} & 	\textbf{1.576} & 	\textbf{-0.933} & -0.390 \\ 
		& 	\textbf{(2.667, 3.480)} & 	\textbf{(0.860, 2.291)} & 	\textbf{(-1.567, -0.298)} & (-1.063, 0.283) \\ 
		\multirow{2}{*}{\emph{A.Dorsalis}} & 0.106 & 0.305 & -0.144 & 	\textbf{2.557} \\ 
		& (-0.513, 0.725) & (-0.544, 1.153) & (-1.064, 0.776) & 	\textbf{(1.720, 3.394)} \\ 
		\multirow{2}{*}{\emph{B.Aeneum}} & 	\textbf{2.010} & 	\textbf{1.467} & 	\textbf{-2.993} & -0.236 \\ 
		& 	\textbf{(1.265, 2.756)} & 	\textbf{(0.262, 2.673)} & 	\textbf{(-4.314, -1.672)} & (-1.345, 0.873) \\ 
		\multirow{2}{*}{\emph{B.Guttula}} & 	\textbf{1.948} & 	\textbf{0.921} & 	\textbf{-1.024} & 0.448 \\ 
		& 	\textbf{(1.560, 2.336)} & 	\textbf{(0.264, 1.577)} & 	\textbf{(-1.672, -0.376)} & (-0.159, 1.055)  \\  \hline
		& &Estimation of $\rho_k$& & \\
		\hline 
		Total Length & Leg Color & Wing Development & Overwintering & Breeding Season \\ 
		\hline
		\textbf{0.061} & -0.002 & 0.021 & 0.004 & 	\textbf{0.036} \\ 
		\textbf{(0.023, 0.098)} & (-0.022, 0.018) & (-0.001, 0.044) & (-0.018, 0.026) & 	\textbf{(0.008, 0.064)} \\  
		\hline
	\end{tabular}}   \vspace{-1em}
		          
	\label{table:carabid_result}                                                                                                                                                  
\end{table}  

Table \ref{table:carabid_result} reports the estimated mean regression coefficients for the first ten carabid ground beetle species, along with the estimated correlation regression parameters for all five trait variables; see supplementary material \ref{sec:supp_ground_beetle} for full estimation results of all $p=38$ species. As expected, carabid ground beetle species possessed considerable heterogeneity in their responses to the environment. For instance, in comparing \emph{A.Muelleri} and \emph{A.Communis}, results show the former preferred higher levels of soil pH, while the latter presented no clear evidence of being influenced by this habitat factor. Although elevation was negatively associated with both species i.e., both tended to be recorded at lower altitude, it was only statistically significant for \emph{A.Muelleri}. Conversely, only \emph{A.Communis} exhibited statistically clear evidence of being negatively impacted by increased land management intensity. Elevation was negatively associated with all the abundances of the first ten carabid species in Table \ref{table:carabid_result}, although we note that 11 other carabid species exhibited positive estimated mean regression coefficients for the elevation covariate (see Table \ref{table:carabid_full_result} in supplementary material \ref{sec:supp_ground_beetle}).

Turning to the estimated correlation regression parameters, all trait variables except leg color exhibited a positive association with the correlation between the abundances of different carabid species. The correlation regression parameter estimate related to total length displayed the strongest magnitude, and its confidence interval excluded zero. This suggested conditional on other trait values, the abundances of carabid species with similar total lengths were more positively associated after accounting for differences in their mean abundances due to environmental filtering. There was also statistically clear evidence that breeding season was important in driving residual covariations between carabid species.

\label{page:sensitivity_analysis_1_start}In supplementary material \ref{sec:supp_ground_beetle}, we performed some additional analyses to investigate the sensitivity of the proposed model to alternative specifications of the similarity measures. Results show the estimated mean regression coefficients of all beetle species, the correlation regression parameters of the trait similarity matrices, and the between-species correlation matrix remained similar across different specifications of similarity measures in this application. This consistency in conclusions further substantiates the inferences drawn from Table \ref{table:carabid_result}.\label{page:sensitivity_analysis_1_end}

\section{Conclusion}\label{sec:conclusion}
We have introduced a joint mean and correlation regression model for correlated multi-response data, which allows simultaneous analysis of the relationship between the mean components with observed covariates, and the association between the correlation components with similarity measures of additional predictor information. The proposed model can be applied to a wide variety of discrete and (semi-)continuous responses. We developed a joint estimator for the mean regression coefficients and correlation regression parameters, which is demonstrated to be consistent and asymptotically normal as the sample size $n\to \infty$ and the number of responses $p$ grows with increasing $n$. \label{page:beta_robust1_start}Similar to standard GEEs, the consistency of the proposed mean regression coefficient estimators is robust to misspecification of the correlation regression component of the joint model.\label{page:beta_robust1_end} \label{page:alternative_method_conclusion_start}Simulation studies demonstrate the strong empirical performance of the proposed joint estimator, especially compared with existing methods such as GEE assuming an unstructured working correlation matrix in estimating the true correlation matrix.\label{page:alternative_method_conclusion_end} The application of our proposed model to the ground beetle dataset revealed heterogeneous relationships between environmental factors and carabid beetle abundances, while also identifying important functional traits which drive the (residual) correlation between species. \label{page:sensitivity_analysis_conclusion_start}A sensitivity analysis also yielded similar conclusions for the same dataset under alternative specifications of similarity measures. \label{page:sensitivity_analysis_conclusion_end}

Note the findings from fitting our proposed model to the ground beetle dataset are novel and differ from those obtained from the fourth corner analysis by \cite{riberaETAL2001}, and indeed more generally fourth corner analyses in community ecology \citep[e.g.,][]{niku2021analyzing}. In particular, while fourth corner models aim at identifying where traits mediate species responses to the environment, our approach quantifies how different traits influence the residual correlation between species after accounting for species-environmental responses. These are fundamentally different scientific questions, and while statistical methods have been developed for the former, our method is one of the first which specifically addresses the latter in statistical ecology.

\label{page:mixed_response_start}A logical next step would be to extend the proposed model to handle mixed response types. This could be achieved by considering link functions $g_{j}(\cdot)$ and variance functions $h_j(\cdot)$ that are allowed to differ based on the $j$-th response type, and Condition \ref{condition:continous} would be altered so that its requirements apply to all link and variance functions. This could be useful in ecology, say, when we observe count records for some species and presence-absence records for other species, in which case $g_{j}(\mu) = \log(\mu)$ and $h_j(\mu;\phi) = \phi \mu$ for the count responses while $g_{j}(\mu) = \log\{ \mu/ (1-\mu) \}$ and $h_j(\mu;\phi) = \phi \mu (1-\mu)$ for the binary responses, noting that there might be complications in the interpretation of the response-specific mean regression coefficients because of the use of different link functions resulting in differing effect sizes \citep[see also Section 6 of][]{huiETAL2024}.\label{page:mixed_response_end} It would also be interesting to consider similarity matrices $\bmW_k^{(i)}$ that are heterogeneous across $i=1,\cdots,n$; this would give rise to heterogeneous correlation matrices $\bmR^{(i)}$  for different response vectors $\bmY_i$ \citep[e.g.,][]{ZOU2021}. \label{page:unbalanced_data_start}Additionally, while this article focuses on a balanced setting by assuming each of the $i$-th cluster consists of $p$ number of responses, the proposed method could be generalized to accommodate unbalanced data where we observe $p_i$ rather than $p$ responses at the $i$-th cluster \citep[analogous to the study of clustered data with unequal cluster sizes e.g., Section 3.5 of][]{xueETAL2010}. In this instance, similar idea of correlation regression could be employed to model the $i$-th cluster correlation matrix as $\bm{R}^{(i)}(\bmrho) = \bm{I}_{p_i} + \sum_{k=1}^{K} \rho_k \bm{W}_k^{(i)}$ involving $p_i \times p_i$ matrices, which could be considered as one special case of the heterogeneous similarity matrices discussed above. The proposed method could then be adapted by appropriately adjusting the dimensions of the matrices involved in the estimation procedures to reflect the varying cluster sizes e.g., $\tilde{\bm{W}}_k$ would be changed from $\bm{I}_n \otimes \bm{W}_k$ under the balanced data setting to $\mathrm{diag}( \bm{W}_k^{(1)}, \cdots, \bm{W}_k^{(n)} )$.\label{page:unbalanced_data_end}

\label{page:spatial_temporal_exetnsion_start}Finally, the proposed model can be extended to allow for the clusters to be spatially and/or temporally correlated across $i=1,\cdots,n$ e.g., by considering the generalized Kronecker product structure of \cite{bonat2016} that involves separate modeling of the between-response and between-cluster dependence structures. In this case, the correlation regression model would still be used to model the between-response dependence structure, while similar idea could be employed to model the between-cluster dependence structure as a linear combination of similarity matrices constructed from spatio-temporal distances between different clusters \citep[see also][for a similar idea]{hui2022gee}.\label{page:spatial_temporal_exetnsion_end} \label{page:similarity_selection_start}In such cases, it may be particularly useful to consider joint variable selection on the mean regression coefficients and correlation regression parameters as the number of covariates $d$ and similarity matrices $K$ could be large in real practice. This could be accomplished by leveraging existing work on penalized GEEs or fast information criterion \citep{wang2012penalized,hui2023gee}; see also the recent work of \cite{thoETAL2024} who considered variable selection to identify the subset of relevant similarity matrices in the context of Ising models.\label{page:similarity_selection_end}  

\section*{Appendix} 
\appendix
We introduce five regularity conditions under which all theoretical results are established. Let $ \|\cdot\|_t $ denote the vector $ t $-norm or matrix $ t $-norm for $ 1 \leq t \leq \infty $, $ |\bm{H}|_\infty = \|\textrm{vec}(\bm{H}) \|_\infty $ denote the element-wise $ \ell_\infty $ norm for a generic matrix $ \bm{H} $ with $ \textrm{vec}(\bm{H}) $ being the vectorization of matrix $ \bm{H} $, and $ \lambda_{\min}(\bm{H}) $ denote the smallest eigenvalue of a generic square matrix $ \bm{H} $. The discussions of the following conditions are presented in supplementary material \ref{sec:condition_discussion}.

\begin{condition}
	The elements $ \varepsilon_{\ell} $ of the $ np $-dimensional random vector $ \bmvarepsilon(\bmvarthetao) = \tilde{\bm{L}}^{-1}_0 \bmANhalf(  \bmbeta^{(0)}) \{\bmY - \bmmu(\bmbeta^{(0)})	\} $ are independent and identically distributed with mean zero, variance one, third-order moment $ \mu^{(3)} $, and fourth-order moment $ \mu^{(4)} $, where $\tilde{\bm{L}}_0 = \bm{I}_n \otimes \bm{L}_0$, and $\bm{L}_0$ is obtained through the Cholesky decomposition of $\bmSigma (\bmalphao) = \bm{L}_0 \bm{L}_0^\top$. Furthermore, there exists some $ \eta > 0 $ such that $ E(| \varepsilon_{\ell} |^{4+\eta}) < \infty $ for $\ell = 1,\cdots,np$.
	\label{condition:varepsilon}
\end{condition}

\begin{condition}
	There exist finite positive constants $C_L, C_X$ and $C_W$ such that $ \max\{ \| \bm{L}_0 \|_1, \| \bm{L}_0^{-1} \|_{1}, \| \bm{L}_0 \|_\infty, \| \bm{L}_0^{-1} \|_\infty \} \leq C_L $,  $  |\bmX |_\infty \leq C_X  $, and $  \| \bmW_k \|_1 \leq C_W    $ for any $n \geq 1$, $p \geq 1$ and $  k = 1,\cdots,K $, where $\bmX = (\bm{x}_1,\cdots,\bm{x}_n)^\top$.
	\label{condition:finite_norm}
\end{condition}

\begin{condition}
	There exists some finite positive constant $C_\beta$ such that $\| \bmbetao \|_\infty \leq C_{\beta}$ for any $p \geq 1$.
	\label{condition:finite_parameters}
\end{condition}
\begin{condition}
	The inverse link function $ g^{-1}(\cdot)$ is twice continuously differentiable, the variance function $ h(\cdot) $  is continuously differentiable, and the composition function $ (h \circ g^{-1})(\cdot) \geq C_h$ for some finite positive constant $C_h$.
	\label{condition:continous}
\end{condition}

\begin{condition}
	There exist finite positive constants $C_{B}$ and $C_{\Omega}$ such that \\ $\max\{  \|\bm{B}^{-1}\|_1,  \|\bm{B}^{-1}\|_\infty\} \leq C_{B}$ and  $\lambda_{\min} ( \bm{\Omega}) \geq C_{\Omega}$ for any $n\geq 1$ and $p \geq 1$, where $\bm{\Omega} = \bm{B}^{-1} \bm{U} \bm{B}^{-1\top}$.
	\label{condition:B_and_Omega}
\end{condition}


	\section*{Supplementary Materials}
	
	The Supplementary Material includes all proofs and algorithms, as well as additional results for the simulation study and real data application.
	\par

\section*{Acknowledgements}
Zhi Yang Tho was supported by an Australian Government Research Training Program scholarship. Francis KC Hui was supported by an Australian Research Council Discovery Project DP240100143. Thanks to Alan Welsh for useful comments.
	

\bibhang=1.7pc
\bibsep=2pt
\fontsize{9}{14pt plus.8pt minus .6pt}\selectfont
\renewcommand\bibname{\large \bf References}
\expandafter\ifx\csname
natexlab\endcsname\relax\def\natexlab#1{#1}\fi
\expandafter\ifx\csname url\endcsname\relax
  \def\url#1{\texttt{#1}}\fi
\expandafter\ifx\csname urlprefix\endcsname\relax\def\urlprefix{URL}\fi

\bibliographystyle{chicago}      
\bibliography{bibliography}   

\vskip .63cm
\noindent
Zhi Yang Tho
\vskip 2pt
\noindent
The Australian National University, Canberra, ACT 2600, Australia.
\vskip 2pt
\noindent
E-mail: zhiyang.tho@anu.edu.au
\vskip 2pt

\noindent
Francis K.C. Hui
\vskip 2pt
\noindent
The Australian National University, Canberra, ACT 2600, Australia.
\vskip 2pt
\noindent
E-mail: francis.hui@anu.edu.au
\vskip 2pt

\noindent
Tao Zou
\vskip 2pt
\noindent
The Australian National University, Canberra, ACT 2600, Australia.
\vskip 2pt
\noindent
E-mail: tao.zou@anu.edu.au

\end{document}